\documentclass[article,aps,nofootinbib,showpacs,amsfonts,epsf]{revtex4}

\input{epsf.tex}

\newcommand{\E}{{\cal{E}}}
\newcommand{\s}{\sigma}

\newcommand{\be}{\begin{equation}}
\newcommand{\ee}{\end{equation}}
\newcommand{\bea}{\begin{eqnarray}}
\newcommand{\eea}{\end{eqnarray}}
\newcommand{\ba}{\begin{array}}
\newcommand{\ea}{\end{array}}
\def\J#1#2#3#4{{#1} {\bf #2}, #3 (#4)}
\def\PRD{Phys. Rev. D}
\def\PR{Phys. Rev.}
\def\PRL{Phys. Rev. Lett.}

\def\JMP{J. Math. Phys.}

\def\CQG{Class. Quantum Grav.}

\def\GRG{Gen. Relativ. Grav.}
\def\PLA{Phys. Lett. A}

\begin{document}
\draft
\title{Metric of a rotating charged magnetized sphere}

\author{V.S. Manko$^a$, I.M. Mej\'ia$^a$, E. Ruiz$^b$}
\address{$^a$Departamento de F\'\i sica, Centro de Investigaci\'on y
de Estudios Avanzados del IPN, A.P. 14-740, 07000 Ciudad de
M\'exico, Mexico\\$^b$Instituto Universitario de F\'{i}sica
Fundamental y Matem\'aticas, Universidad de Salamanca, 37008
Salamanca, Spain}

\begin{abstract}
Stationary axisymmetric metric describing the exterior field of a
rotating, charged sphere endowed with magnetic dipole moment is
presented and discussed. It has a remarkably simple multipole
structure defined by only four nonzero Hoenselaers-Perj\'es
relativistic moments.
\end{abstract}

\pacs{04.20.Jb, 04.70.Bw, 97.60.Lf}

\maketitle

\section{Introduction}

In 1984, Simon \cite{Sim} proposed a definition of the
relativistic multipole moments for the stationary axisymmetric
electrovacuum spacetimes generalizing the known Geroch-Hansen
definition \cite{Ger,Han} developed for the stationary vacuum
case. The practical computation of Simon's multipoles is
facilitated by the Hoenselaers-Perj\'es (HP) procedure \cite{HPe},
rectified by Sotiriou and Apostolatos \cite{SAp}, according to
which these multipoles are expressible in terms of the
coefficients in the series expansions of the modified Ernst
potentials \cite{Ern} $\xi$ and $\eta$ evaluated on the symmetry
axis and related to the usual Ernst potentials $\E$ and $\Phi$ by
the formulas
\be \xi=\frac{1-\E}{1+\E}, \quad \eta=\frac{2\Phi}{1+\E}.
\label{xi} \ee
In the papers on exact electrovacuum solutions it is customary to
calculate the first four Simon's complex moments $P_n$ and $Q_n$,
$n=0,1,2,3$, which coincide with the corresponding HP coefficients
$m_n$ and $q_n$ in the expansions ($z\to\infty$)
\be \xi(\rho=0,z)= \sum\limits_{n=0}^{\infty}m_nz^{-n-1}, \quad
\eta(\rho=0,z)= \sum\limits_{n=0}^{\infty}q_nz^{-n-1},
\label{mnqn} \ee
because they provide one with important information about the
physical characteristics of the sources such as the total mass,
angular momentum, electric charge or magnetic dipole moment. For
all practical applications, the knowledge of Simon's moments
higher than $n=3$ is not actually needed, and in this respect it
would be worthy to note that even the expressions of $P_n$ and
$Q_n$ for $n=4$ and 5 were given in \cite{HPe} with errors
detected only fourteen years later \cite{SAp}.

In our recent work \cite{MMR} we have introduced the notion of the
Fodor-Hoenselaers-Perj\'es (FHP) multipole moments \cite{FHP} for
vacuum spacetimes as an alternative to the Geroch-Hansen (GH)
multipoles \cite{Ger,Han}. The objective of the present paper is
to extend our previous results to the electrovacuum case and in
particular give arguments in favor of introducing the {\it HP
multipole moments} instead of the Simon's ones. As a nontrivial
example of a spacetime determined by only four nonzero HP moments
we shall consider an electrovac metric for a spinning sphere
endowed with electric charge and magnetic dipole moment.

\section{The 5-parameter solution and its 4-parameter subfamily}

As has been observed in the paper \cite{MMR}, it is the GH
multipole moments $P_n$ that should be considered approximations
to the FHP quantities $m_n$, and not the contrary. This is because
the knowledge of the axis value of the Ernst potential, which is
uniquely determined by the FHP multipoles $m_n$, is sufficient for
its holomorphic continuation to the whole space and for
calculating the corresponding metric functions, and the GH moments
$P_n$ are completely dropped from such a modern solution
generating procedure. At the same time, it is also clear that the
quantities $P_n$ and $m_n$ are rather closely related, the first
four of them being identical, $P_n=m_n$, $n=0,1,2,3$, while in the
particular case of the Kerr metric \cite{Ker} the latter equality
holds for all $n$. The particular 3-parameter solution for a
spinning deformed mass considered in \cite{MMR} is defined by the
axis data
\be \E(\rho=0,z)=\frac{z^2-Mz-M^2q-iM^2j}{z^2+Mz-M^2q+iM^2j},
\label{axis3} \ee
where $M$, $q$ and $j$ are, respectively, the mass, dimensionless
mass quadrupole moment and dimensionless angular momentum. In the
limit $q=0$, the axis data (\ref{axis3}) defines the solution
describing the exterior field of a rotating sphere because in this
case all the corresponding FHP mass multipole moments, except for
the monopole one, become equal to zero.

We find it likely, in view of the potential interest for physical
and astrophysical applications, to generalize the results of the
paper \cite{MMR} to the electrovacuum case by introducing the
additional parameters of electric charge and magnetic dipole
moment. Then we must consider the axis data of the form
\bea \E(\rho=0,z)=\frac{z^2-Mz-M^2q-iM^2j}{z^2+Mz-M^2q+iM^2j},
\nonumber\\ \Phi(\rho=0,z)=\frac{Mez+iM^2\mu}{z^2+Mz-M^2q+iM^2j},
\label{axis5} \eea
where the electromagnetic field is described by the Ernst
potential $\Phi$ in which the parameters $e$ and $\mu$ are the
dimensionless charge and the dimensionless magnetic dipole moment,
respectively. The latter interpretation can be confirmed by
calculating the first four complex multipole moments with the aid
of formulas (\ref{mnqn}), thus yielding for (\ref{axis5})
\bea m_0=M, \quad m_1=iM^2j, \quad m_2=M^3q, \quad m_3=iM^4qj, \nonumber\\
q_0=Me, \quad q_1=iM^2\mu, \quad q_2=M^3qe, \quad q_3=iM^4q\mu,
\label{mq4} \eea
and the above HP multipoles coincide with the respective Simon's
multipole moments. As will be seen below, the explicit formulas
for the coefficients $m_n$ and $q_n$ determined by the data
(\ref{axis5}) can be readily found for any $n$, whereas the
general expressions of the Simon multipoles $P_n$ and $Q_n$ for
$n>5$ have not been computed to date because of their complicated
form and scarce significance. Therefore, taking into account that
the knowledge of the axis data (\ref{axis5}) is sufficient for the
construction of the corresponding Ernst potentials in the entire
space \cite{Sib}, and also that in general the axis values of the
Ernst potentials are defined uniquely by the quantities $m_n$ and
$q_n$, it is natural to come to the conclusion that Simon's
multipoles $P_n$ and $Q_n$ must make room for what we can rightly
call the Hoenselaers-Perj\'es multipole moments $m_n$ and $q_n$
which are better adjusted to the intrinsic structure of the
stationary electrovacuum solutions and to the modern solution
generating techniques.

The general concise expressions for the HP multipole moments of
the 5-parameter solution defined by the data (\ref{axis5}) can be
shown to have the form
\bea &&m_{2k}=M^{2k+1}q^k, \quad m_{2k+1}=iM^{2k+2}q^kj, \nonumber\\
&&q_{2k}=M^{2k+1}q^ke, \quad q_{2k+1}=iM^{2k+2}q^k\mu, \quad
k=0,1,2,\ldots \label{mqn} \eea
and the Ernst potentials possessing the above multipole structure
can be constructed from (\ref{axis5}) by means of Sibgatullin's
integral method \cite{Sib,MSi}. The resulting expressions for $\E$
and $\Phi$ are
\bea \E&=&\frac{A-B}{A+B}, \quad \Phi=\frac{C}{A+B}, \nonumber\\
A&=&\s_+\s_-[(\s_+^2+\s_-^2)(1-e^2)+2(j^2-\mu^2)]
(R_++R_-)(r_++r_-) \nonumber\\ &&-[2\s_+^2\s_-^2(1-e^2)
+(\s_+^2+\s_-^2)(j^2-\mu^2)](R_+-R_-)(r_+-r_-) \nonumber\\
&&+4\s_+\s_-(j^2+q-qe^2-\mu^2)(R_+R_-+r_+r_-) \nonumber\\
&&+id(j-e\mu)[\s_+(R_++R_-)(r_+-r_-) -\s_-(R_+-R_-)(r_++r_-)],
\nonumber\\
B&=&Md\{\s_+\s_-[d(R_++R_-+r_++r_-) -(1-e^2)(R_++R_--r_+-r_-)]
\nonumber\\ &&-i(j+je^2-2e\mu)[\s_-(R_+-R_-)-\s_+(r_+-r_-)]
\nonumber\\ &&+ijd[\s_-(R_+-R_-)+\s_+(r_+-r_-)]\},  \nonumber\\
C&=&Md\{e\s_+\s_-[d(R_++R_-+r_++r_-) -(1-e^2)(R_++R_--r_+-r_-)]
\nonumber\\ &&-i(2je-\mu-e^2\mu)[\s_-(R_+-R_-)-\s_+(r_+-r_-)]
\nonumber\\ &&+i\mu d[\s_-(R_+-R_-)+\s_+(r_+-r_-)]\},  \nonumber\\
R_\pm&=&\sqrt{\rho^2+\left(z\pm M\s_+\right)^2}, \quad
r_\pm=\sqrt{\rho^2+\left(z\pm M\s_-\right)^2},
\nonumber\\
\s_\pm&=&\sqrt{(1+2q-e^2\pm d)/2}, \nonumber\\
d&=&\sqrt{(1+2q-e^2)^2+4(j^2-q^2-\mu^2)}, \label{EF5} \eea
and these have been worked out with the aid of the general
formulas of the paper \cite{MMR2}.

The well-known Kerr-Newman (KN) solution \cite{KNe} for a charged
rotating mass is contained in (\ref{EF5}) as the particular case
$q=-j^2$, $\mu=je$, for which we get from (\ref{mqn})
\be m_{n}=M(iMj)^{n}, \quad q_{n}=Me(iMj)^{n}, \quad
n=0,1,2,\ldots \label{mqKN} \ee
and these $m_n$ and $q_n$ coincide with the Simon multipoles $P_n$
and $Q_n$ calculated for the KN solution by Sotiriou and
Apostolatos \cite{SAp}. Apparently, the KN spacetime is determined
by an infinite set of multipole moments. At the same time, as it
follows from (\ref{mq4}) and (\ref{mqn}), the solution (\ref{EF5})
has a very interesting 4-parameter subfamily defined by only four
nonzero HP multipole moments which corresponds to the choice $q=0$
in (\ref{EF5}). Since all the mass-multipole moments, except for
the monopole one, in this case are equal to zero, the resulting
solution should be interpreted as describing the exterior geometry
of a rotating charged magnetized sphere. Anticipating a possible
wide interest this 4-parameter electrovac solution might represent
to the researchers due to its remarkable multipole structure, in
what follows we shall consider it in more detail. First of all, we
note that in the case of vanishing $q$ the Ernst potentials
(\ref{EF5}) take the form
\bea \E&=&\frac{A-B}{A+B}, \quad \Phi=\frac{C}{A+B}, \nonumber\\
A&=&\s_+\s_-[(1-e^2)^2+2(j^2-\mu^2)] (R_++R_-)(r_++r_-) \nonumber\\
&&+(j^2-\mu^2)[(1-e^2)(R_+-R_-)(r_+-r_-)
+4\s_+\s_-(R_+R_-+r_+r_-)] \nonumber\\
&&+id(j-e\mu)[\s_+(R_++R_-)(r_+-r_-) -\s_-(R_+-R_-)(r_++r_-)],
\nonumber\\
B&=&Md\{\s_+\s_-[d(R_++R_-+r_++r_-) -(1-e^2)(R_++R_--r_+-r_-)]
\nonumber\\ &&-i\s_-(j+je^2-2e\mu-jd)(R_+-R_-) \nonumber\\
&&+i\s_+(j+je^2-2e\mu+jd)(r_+-r_-)\},  \nonumber\\
C&=&eB-iMd (je-\mu)[\s_-(1-e^2+d)(R_+-R_-) \nonumber\\
&&-\s_+(1-e^2-d)(r_+-r_-)], \nonumber\\
R_\pm&=&\sqrt{\rho^2+\left(z\pm M\s_+\right)^2}, \quad
r_\pm=\sqrt{\rho^2+\left(z\pm M\s_-\right)^2},
\nonumber\\
\s_\pm&=&\sqrt{(1-e^2\pm d)/2}, \quad
d=\sqrt{(1-e^2)^2+4(j^2-\mu^2)}, \label{EF4} \eea
while the corresponding metric functions $f$, $\gamma$ and
$\omega$, which enter the line element
\be d s^2=f^{-1}[e^{2\gamma}(d\rho^2+d z^2)+\rho^2 d\varphi^2]-f(d
t-\omega d\varphi)^2, \label{pap} \ee
can be worked out from the respective general expressions of the
paper \cite{MMR2}, yielding
\bea f&=&\frac{A\bar A-B\bar B+C\bar C}{(A+B)(\bar A+\bar B)},
\quad e^{2\gamma}=\frac{A\bar A-B\bar B+C\bar C}{16d^4|\s_+|^2
|\s_-|^2R_+R_-r_+r_-}, \nonumber\\ \omega&=&-\frac{{\rm Im}
[G(\bar A+\bar B)+C\bar I]}{A\bar A-B\bar B+C\bar C}, \nonumber\\
G&=&2(iMj-z)B+Md\{\s_+\s_-(2-e^2)[\s_-(R_++R_-)(r_+-r_-)\nonumber\\
&&-\s_+(R_+-R_-)(r_++r_-)]-(2j^2-\mu^2)[\s_-(R_+-R_-)(r_++r_-) \nonumber\\
&&-\s_+(R_++R_-)(r_+-r_-)]-id(2j-e\mu)(R_+-R_-)(r_+-r_-)\}
\nonumber\\
&&-iMe(je-\mu)[(1-e^2)(R_+-R_-)(r_+-r_-) \nonumber\\
&&-2\s_+\s_-(R_++R_-)(r_++r_-) +4\s_+\s_-(R_+R_-+r_+r_-)]
\nonumber\\
&&+2M^2d\{2j^2\s_+\s_-[\s_+(R_+-R_-)-\s_-(r_+-r_-)] \nonumber\\
&&+\mu(je-\mu) [\s_-(R_+-R_-)-\s_+(r_+-r_-)] \nonumber\\
&&-ie\s_+\s_-(je-\mu)
(R_++R_--r_+-r_-)\}, \nonumber\\
I&=&-zC+Md[\s_+(e\s_-^2+j\mu)(R_++R_-)(r_+-r_-)
-\s_-(e\s_+^2+j\mu) \nonumber\\
&&\times(R_+-R_-)(r_++r_-)-2i\s_+\s_-(je-\mu)(R_+R_--r_+r_-
+2M^2d)] \nonumber\\
&& -iM\{\s_+\s_-(1+e^2)(je-\mu)[2(R_+R_-+r_+r_-)
-(R_++R_-)(r_++r_-)]
\nonumber\\
&&+[2(je+\mu)(j^2-\mu^2)+e(1-e^2)(j-e\mu)](R_+-R_-)(r_+-r_-)\}
\nonumber\\
&&+M^2d\{2(j\mu+e\mu^2-2j^2e)[\s_+(r_+-r_-)-\s_-(R_+-R_-)]
\nonumber\\
&&+i\s_+\s_-[\mu d(R_++R_-+r_++r_-)- (4je-3\mu-e^2\mu) \nonumber\\
&&\times(R_++R_--r_+-r_-)]\}. \label{mf4} \eea
Moreover, the nonzero electric and magnetic components of the
electromagnetic 4-potential are given by the formulas
\be A_t=-{\rm Re}\left(\frac{C}{A+B}\right), \quad A_\varphi={\rm
Im}\left(\frac{I}{A+B}\right), \label{Atf} \ee
so that we have fully described the gravitational and
electromagnetic fields of the rotating sphere carrying both the
electric charge and magnetic dipole moment.

Evidently, due to its finite multipole structure, the 4-parameter
solution (\ref{EF4})-(\ref{Atf}) does not contain the Kerr and KN
spacetimes as particular cases, and therefore it differs from the
4-parameter solution describing the magnetized KN source
\cite{Man,Man2}. However, in the limit of zero angular momentum
($j=0$), both solutions coincide, representing a magnetized
Reissner-Nordstr\"om mass. It is, therefore, the nonzero angular
momentum sector of the solution (\ref{EF5}) that makes it very
special and attractive from the physical point of view.

\section{Discussion}

It is easy to see that formulas (\ref{EF4}) and (\ref{mf4})
considerably simplify if the parameters $j$, $e$ and $\mu$ are
subject to the constraint $\mu=ej$ which represents the same
gyromagnetic ratio of the electron as in the KN metric (see e.g.
\cite{PDu} for a discussion of this ratio in the context of exact
electrovac solutions). In that subfamily, the potential $\E$
becomes an analytic function of $\Phi$ and hence such a
3-parameter subfamily can be treated within the framework of the
well-known Ernst-Harrison (EH) charging transformation
\cite{Ern,Har} involving the nonzero parameter $e$. Since in the
known practical applications of the EH transformation the values
of $e$ are usually restricted to the ``undercharged'' case $e<1$,
it would be instructive to see how this transformation works in
the ``overcharged'' case $e>1$ too. With this idea in mind, we
first set $\mu=ej$ in the formulas (\ref{EF4}) and (\ref{mf4}),
and then rewrite the resulting solution by rescaling the
quantities $\s_\pm$ and introducing
\be m=M\varepsilon, \quad {\tt j}=j/\varepsilon, \quad
\varepsilon\equiv\sqrt{1-e^2}, \label{mj} \ee
thus finally yielding for the Ernst potentials $\E$ and $\Phi$ the
expressions
\bea \E&=&\frac{\varepsilon A-B}{\varepsilon A+B}, \quad
\Phi=\frac{eB}{\varepsilon A+B}, \nonumber\\
A&=&\s_+\s_-(1+2{\tt j}^2)(R_++R_-)(r_++r_-) +{\tt
j}^2[(R_+-R_-)(r_+-r_-) \nonumber\\
&&+4\s_+\s_-(R_+R_-+r_+r_-)]+i{\tt j}\delta[\s_+(R_++R_-)(r_+-r_-)
\nonumber\\
&&-\s_-(R_+-R_-)(r_++r_-)],
\nonumber\\
B&=&m\delta\{\s_+\s_-[(\delta-1)(R_++R_-)+(\delta+1)(r_++r_-)]
\nonumber\\ &&+i{\tt j}[\s_-(\delta-1)(R_+-R_-)
+\s_+(\delta+1)(r_+-r_-)]\}, \nonumber\\
R_\pm&=&\sqrt{\rho^2+\left(z\pm m\s_+\right)^2}, \quad
r_\pm=\sqrt{\rho^2+\left(z\pm m\s_-\right)^2},
\nonumber\\
\s_\pm&=&\sqrt{(1\pm \delta)/2}, \quad \delta=\sqrt{1+4{\tt j}^2},
\label{EF3} \eea
and for the metric functions $f$, $\gamma$, $\omega$ the
expressions
\bea f&=&\frac{\varepsilon\bar\varepsilon A\bar A
-\varepsilon^2B\bar B} {(\varepsilon A+B)(\bar\varepsilon\bar
A+\bar B)}, \quad e^{2\gamma}=\frac{\varepsilon\bar\varepsilon
A\bar A -\varepsilon^2B\bar B}{16\delta^4|\varepsilon|^2|\s_+|^2
|\s_-|^2
R_+R_-r_+r_-}, \nonumber\\
\omega&=&-\frac{{\rm Im} [G(\bar\varepsilon\bar A+\bar B)+eB\bar
I]}{\varepsilon\bar\varepsilon A\bar A
-\varepsilon^2B\bar B}, \nonumber\\
G&=&2(im{\tt j}-z)B+\frac{1+\varepsilon^2}{\varepsilon}
m\delta\{\s_+\s_-[\s_-(R_++R_-)(r_+-r_-) \nonumber\\
&&-\s_+(R_+-R_-)(r_++r_-)]-{\tt j}^2[\s_-(R_+-R_-)(r_++r_-) \nonumber\\
&&-\s_+(R_++R_-)(r_+-r_-)]-i{\tt j}\delta(R_+-R_-)(r_+-r_-)\}
\nonumber\\
&&+4m^2{\tt j}^2\delta\s_+\s_-[\s_+(R_+-R_-)-\s_-(r_+-r_-)], \nonumber\\
I&=&-ezB+\frac{e}{\varepsilon}m\delta[\s_+(\s_-^2+{\tt
j}^2)(R_++R_-)(r_+-r_-) \nonumber\\ &&-\s_-(\s_+^2+{\tt
j}^2)(R_+-R_-)(r_++r_-) -i{\tt j}\delta(R_+-R_-)(r_+-r_-)]
\nonumber\\
&&+m^2{\tt j}e\delta\{2{\tt j}[\s_-(R_+-R_-)-\s_+(r_+-r_-)]
\nonumber\\
&&+i\s_+\s_-[(\delta-1)(R_++R_-)+(\delta+1)(r_++r_-)]\}.
\label{mf3} \eea

Formulas (\ref{mj}) and (\ref{EF3}) determine the essence of the
EH transformation: this symmetry transformation generates an
electrovac solution from a given vacuum one by introducing the
charge parameter $e$ by means of the relations (\ref{EF3}) for
$\E$ and $\Phi$. At the same time, as it follows from (\ref{mj}),
while in the undercharged case ($e<1$) the rescaled parameters $m$
and ${\tt j}$ remain real-valued, the latter parameters become
pure imaginary in the overcharged case $e>1$, which, however, does
not mean that the branch $e>1$ of the EH transformation is
unphysical -- one simply has to take into account the relation of
the parameters $m$ and ${\tt j}$ to the physical mass $M$ and
physical angular momentum $j$ given in (\ref{mj}). Mention also
that in the overcharged case the expression for the metric
function $\gamma$, as can be seen in (\ref{mf3}), is not of the
same form as in the vacuum solution because $\varepsilon$ is pure
imaginary for $e>1$. All these subtleties may explain why the
application of the EH transformation is restricted in the
literature to the simpler undercharged case $e<1$ only.

Another interesting subclass of the solution
(\ref{EF4})-(\ref{mf4}) is defined by the condition of vanishing
electric charge $e=0$. The resulting 3-parameter metric could be
interpreted as a rotating magnetized sphere, thus being
appropriate for the description of the exterior field of a neutron
star with negligible deformation. The expressions of the
potentials $\E$ and $\Phi$ in this case take the form
\bea \E&=&\frac{A-B}{A+B}, \quad \Phi=\frac{i\mu C}{A+B}, \nonumber\\
A&=&\s_+\s_-(1+2j^2-2\mu^2) (R_++R_-)(r_++r_-) \nonumber\\
&&+(j^2-\mu^2)[(R_+-R_-)(r_+-r_-)
+4\s_+\s_-(R_+R_-+r_+r_-)] \nonumber\\
&&+ijd[\s_+(R_++R_-)(r_+-r_-) -\s_-(R_+-R_-)(r_++r_-)],
\nonumber\\
B&=&Md\{\s_+\s_-[(d-1)(R_++R_-)+(d+1)(r_++r_-)]
\nonumber\\ &&+ij[\s_-(d-1)(R_+-R_-) +\s_+(d+1)(r_+-r_-)]\},  \nonumber\\
C&=&Md[\s_-(d+1)(R_+-R_-) +\s_+(d-1)(r_+-r_-)], \nonumber\\
R_\pm&=&\sqrt{\rho^2+\left(z\pm M\s_+\right)^2}, \quad
r_\pm=\sqrt{\rho^2+\left(z\pm M\s_-\right)^2},
\nonumber\\
\s_\pm&=&\sqrt{(1\pm d)/2}, \quad d=\sqrt{1+4(j^2-\mu^2)},
\label{EF3m} \eea
whereas for the corresponding metric coefficients we readily get
\bea f&=&\frac{A\bar A-B\bar B+\mu^2C\bar C}{(A+B)(\bar A+\bar
B)}, \quad e^{2\gamma}=\frac{A\bar A-B\bar B+\mu^2C\bar
C}{16d^4|\s_+|^2 |\s_-|^2R_+R_-r_+r_-}, \nonumber\\
\omega&=&-\frac{{\rm Im}
[G(\bar A+\bar B)+\mu^2C\bar I]}{A\bar A-B\bar B+\mu^2C\bar C}, \nonumber\\
G&=&2(iMj-z)B+Md\{2\s_+\s_-[\s_-(R_++R_-)(r_+-r_-)\nonumber\\
&&-\s_+(R_+-R_-)(r_++r_-)]-(2j^2-\mu^2)[\s_-(R_+-R_-)(r_++r_-) \nonumber\\
&&-\s_+(R_++R_-)(r_+-r_-)]-2ijd(R_+-R_-)(r_+-r_-)\}
\nonumber\\
&&+2M^2d\{2j^2\s_+\s_-[\s_+(R_+-R_-)-\s_-(r_+-r_-)] \nonumber\\
&&-\mu^2 [\s_-(R_+-R_-)-\s_+(r_+-r_-)]\}, \nonumber\\
I&=&-zC+M\{\s_+\s_-[2(R_+R_-+r_+r_-) -(R_++R_-)(r_++r_-)]
\nonumber\\
&&-2(j^2-\mu^2)(R_+-R_-)(r_+-r_-)\}
\nonumber\\
&&+Md\{2\s_+\s_-(R_+R_--r_+r_- +2M^2d) \nonumber\\
&&+ij[\s_-(R_+-R_-)(r_++r_-)-\s_+(R_++R_-)(r_+-r_-)]\}
\nonumber\\
&&+M^2d\{\s_+\s_-[(d+3)(R_++R_-)+(d-3)(r_++r_-)]
\nonumber\\
&&+2ij[\s_-(R_+-R_-)-\s_+(r_+-r_-)]\}. \label{mf3m} \eea
Note that in the above formulas (\ref{EF3m}) and (\ref{mf3m}) the
functions $C$ and $I$ have been slightly redefined compared with
$C$ and $I$ in (\ref{EF4}) and (\ref{mf4}), in order to make more
visual the appearance of the factor $i\mu$ in the zero charge
case. With such redefinitions, the expressions (\ref{Atf}) for
$A_t$ and $A_\varphi$ also slightly change, namely,
\be A_t=-{\rm Re}\left(\frac{i\mu C}{A+B}\right), \quad
A_\varphi={\rm Im}\left(\frac{i\mu I}{A+B}\right). \label{Atf3m}
\ee

Apparently, although the solution for a rotating magnetized
sphere, like the solution (\ref{EF3})-(\ref{mf3}) before, is a
3-parameter specialization of the general metric
(\ref{EF4})-(\ref{mf4}), it has only three nonzero HP multipole
moments, while in the previous example the three parameters define
four nonzero moments. Since both electrovac solutions considered
in this section have the same pure vacuum limit determined by two
gravitational multipoles, representing mass and angular momentum,
we would like to briefly comment in conclusion on an old
misleading statement, originally made in \cite{BSi} and readily
adopted by various authors as a true one, according to which {\it
any stationary, asymptotically flat solution to Einstein's
equation approaches asymptotically the Kerr solution}.
Intuitively, the idea of this statement might look plausible at
first glance, as for instance any stationary axisymmetric
asymptotically flat solution with nonzero mass and nonzero total
angular momentum would indeed have the same leading mass and
rotational moments as in the Kerr solution. However, a simple
counterexample to the above statement is a {\it stationary}
solution for two counter-rotating Kerr sources \cite{MRR} with
nonzero mass and zero total angular momentum because the Kerr
solution in absence of the angular momentum reduces to the {\it
static} Schwarzschild spacetime. Moreover, the case of zero total
mass which was remarked to be also suitably treated in \cite{BSi}
would have nothing to do with the Kerr solution since the limit
$M=0$ in the latter is just the Minkowski space, whereas there is
an infinite number of stationary asymptotically flat spacetimes
with zero total mass.

\section*{Acknowledgments}

This work was partially supported by the CONACYT of Mexico, by
Project PGC2018-096038-B-100 from Ministerio de Ciencia,
Innovaci\'on y Universidades of Spain, and by Project SA083P17
from Junta de Castilla y Le\'on of Spain.

\end{document}